\documentclass[twocolumn]{aastex631}

\newcommand\tm[1]{\textcolor{black}{#1}}
\newcommand\rs[1]{\textcolor{black}{#1}}

\newcommand\old[1]{\textcolor{black}{#1}}

\received{July 2, 2023}
\accepted{\today}

\submitjournal{ApJ}

\usepackage[normalem]{ulem}
\usepackage{threeparttable}
\usepackage{amsmath}

\shorttitle{CSM formation by binary interaction}
\shortauthors{Matsuoka \& Sawada}

\graphicspath{{./}{figures/}}

\begin{document}

\title{{Binary Interaction Can Yield a Diversity of Circumstellar Media around Type II Supernova Progenitors}}

\author[0000-0002-6916-3559]{Tomoki Matsuoka}
\correspondingauthor{Tomoki Matsuoka}
\email{matsuoka.tomoki.0124@gmail.com}
\affiliation{Institute of Astronomy and Astrophysics, Academia Sinica, No.1, Sec. 4, Roosevelt Road, Taipei 10617, Taiwan, R.O.C.}
\affiliation{Department of Earth Science and Astronomy, Graduate School of Arts and Sciences, The University of Tokyo, Tokyo 153-8902, Japan}

\author[0000-0003-4876-5996]{Ryo Sawada}
\affiliation{Department of Earth Science and Astronomy, Graduate School of Arts and Sciences, The University of Tokyo, Tokyo 153-8902, Japan}

\begin{abstract}
Recent observations of supernovae (SNe) have indicated that a fraction of massive stars possess dense circumstellar medium (CSM) at the moment of their core collapses.
They suggest the presence of additional activities of the SN progenitor driving the enhancement of the mass-loss rate, and some physical processes attributing to single star's activities have been considered.
\old{In this study, we carry out binary evolutionary simulations of massive stars by {\tt MESA} and investigate effects on the subsequent CSM formation through hydrodynamical simulations by {\tt PLUTO}.}
We show that the mass-transfer rate in a binary can increase at the beginning of the Roche lobe overflow, and this enhancement would be associated with the structure of the CSM before the explosion.
We also illustrate that depending on the orbital period of the binary, the density structure of the CSM can have a diverse distribution including shell-like and cliff-like structures.
These characteristic structures appear within the lengthscale of $\sim 10^{17}\,{\rm cm}$ and could be traced by long-term observations of SNe, if the slow velocity of the CSM is assumed ($\sim 10\,{\rm km}\,{\rm s}^{-1}$).
Our results highlight the importance of binary interaction in the aspect of reproducing the diversity of the CSM configuration.
\end{abstract}

\keywords{}


\section{Introduction} \label{sec:intro}

Mass loss from massive stars is one of the important elements in their stellar evolution {so that it has} a significant influence on the evolutionary characteristics and the fates of themselves \citep[e.g.,][]{2015PASA...32...15Y}. Understanding the physical properties of mass loss from massive stars would be a clue toward completing the construction of stellar evolution theory. 
Particularly, mass-loss activities prior to core-collapse supernovae (SNe) would be imprinted in the physical property of the circumstellar medium (CSM). This can contribute to the radiative source in SNe \citep[see e.g.,][]{2012ApJ...744...10K, 2017hsn..book..403S, 2017hsn..book..875C}, and thus understanding the configuration of the CSM can also promote the precise modeling of SNe.   

Thanks to the recent development of transient surveys and rapid follow-up observation systems, it has been recognized that there is a diversity of the CSM configuration inferred for SN progenitors. For instance, it is implicated through optical flash spectroscopy experiments that a Type II SN progenitor is encompassed by a dense CSM confined within small lengthscale \citep[$\sim 10^{15}\,{\rm cm}$. See e.g., SN 2013fs, SN 2023ixf,][]{2017NatPh..13..510Y, 2023arXiv230604721J, 2023arXiv231010727Z}. This is also supported by theoretical modelings of the early-phase optical light curve \citep{2018NatAs...2..808F, 2017ApJ...838...28M, 2017MNRAS.469L.108M}, as well as the detection of precursor activities before the SN \citep[e.g., SN 2020tlf,][]{2022ApJ...924...15J}. Another example is that a few SNe are considered to undergo CSM interaction in the late phase \citep[$\sim$ several years after the explosion. See e.g.,][]{2020ApJ...900...11W, 2021MNRAS.504.2073K, 2022ApJ...936..111K}. As an extreme example of the late-phase CSM interaction, it is argued that there are several objects exhibiting the transformation of the SN type from a stripped-envelope SN to a Type IIn SN \citep{2017ApJ...835..140M, 2019ApJ...887...75T, 2021A&A...650A.174T, 2020ApJ...902...55C}, or showing the rebrightening of radio emission \citep{2018MNRAS.478.5050M, 2021ApJ...923...32B, 2023ApJ...945L...3M}. These objects are speculated to have shell-structured CSM detached from the progenitor, highlighting the possible mass ejection of the progenitor in the past \citep[see the discussion in][]{2023ApJ...945L...3M}.

The CSM structure deviating from the steady wind configuration is thought to attribute to an enhanced mass-loss activity before the explosion. 
Since canonical stellar evolution theories do not expect a significant increase in the mass-loss rate before the SN explosion \citep{2012ARA&A..50..107L}, several mechanisms attempting the explanation of the increase in the mass-loss rate have been proposed, including the excitation of gravity wave in the convective core \citep[e.g.,][]{2012MNRAS.423L..92Q, 2014ApJ...780...96S,2014ApJ...785...82S, 2017MNRAS.470.1642F, 2020ApJ...891L..32M, 2021ApJ...906....3W,2022ApJ...930..168K}. 
It should be noted that these attempts have focused on the properties of the stellar evolutionary behavior as a single star. 

In this study, we simulate the stellar evolution in a binary system to calculate the pre-SN mass-loss rate of massive stars and propose the scenario that binary interaction can enhance the mass-loss rate from the binary system before the explosion.
It is widely known that most massive stars are involved in a binary system that is about to undergo binary interaction \citep{2012Sci...337..444S}.
As the star evolves with time it experiences the expansion or contraction of the stellar radius at the moment of switching of the nuclear burning phases. If the star is involved in a tight binary so that the mass transfer between the stars can initiate, the mass exchange or the escape from the binary can affect the subsequent evolution of the primary star.
In addition, the gas expelled from the binary system can be distributed around the SN progenitor as a CSM.
Simulations of stellar evolution in binary systems have been examined in previous studies \citep[e.g.,][]{2017ApJ...840...90O,2020A&A...637A...6L,2021A&A...656A..58L}, but none of them have discussed CSM formation based on the mass-loss histories obtained therein.
Through the demonstrations in this paper, we illustrate the variety of the CSM molded by binary interaction and suggest the \old{possibility that this variety can give rise to the observational features of subsequent SNe seen a few years after the explosions. Our results shed light on the} necessity to take into consideration the contribution of binary interaction to the formation of the CSM.

\old{After the submission of this paper, we found that \citet{2023arXiv230801819E} conducted a series of stellar evolutionary simulations to systematically survey the properties of long-period binaries of massive stars, which may largely overlap with the content of this paper.
However, one of the pioneering points of this study is that we, for the first time, associate the activity of binary interaction with the formation of the CSM, which is enthusiastically investigated with the tool of hydrodynamical simulations.}
\old{Another is that we attempt to give interpretations on time-variable mass-loss activities of SN progenitors in the framework of binary interaction.}

This paper is organized as follows. In Section \ref{sec:mam} the setup and procedure of our simulations are described. In Section \ref{sec:result} we show that for binary systems with specific parameter sets, a significant increase in the mass-loss rate before the SN explosion can be expected. We also demonstrate that the synthesized mass-loss history model can reproduce the inhomogeneous CSM structure deviating from the steady wind. In Section \ref{sec:discussion} we discuss \old{possible physical processes that can alter our results. Especially, the dependence of the primary star's initial mass is quantitatively investigated.} Finally, we summarize the content of this paper in Section \ref{sec:summary}.

\section{Models and Method} \label{sec:mam}
Our study can be divided into two main parts: the estimation of the mass loss rate by binary evolution calculations using {\tt MESA}, and the reconstruction of the CSM from this mass-loss history by hydrodynamic calculations using {\tt PLUTO}.
We show the setup of {\tt MESA} for the binary evolution calculation in Section \ref{subsec:mesa}, and the results of the mass-loss rate estimation in Section \ref{subsec:binary}. Similarly, the setup of the hydrodynamic calculations in {\tt PLUTO} is shown in Section \ref{subsec:pluto}, and the results of the reconstruction of the CSM structure are in Section \ref{subsec:csm}.

\subsection{Setup of Stellar Evolution code {\tt MESA}}\label{subsec:mesa}
We use stellar evolution code {\tt MESA} in revision {\tt15140}  \citep{2011ApJS..192....3P,2013ApJS..208....4P,2015ApJS..220...15P,2018ApJS..234...34P,2019ApJS..243...10P} to solve stellar evolutions of both two stars in the binary
from the zero-age main sequence (ZAMS).
The computational setup and input parameters in {\tt MESA} are mainly based on \citet{2017ApJ...840...90O}, with some references to \citet{2010ApJ...725..940Y,2017ApJ...840...10Y} and \citet{2020A&A...637A...6L,2021A&A...656A..58L}.
{Hereafter, we describe below the notable parts of this study.}

We start the stellar evolutionary simulations with a non-rotating star for all the models. 
The initial ZAMS mass of the primary star is set to $M_1=12M_\odot$, which is a typical mass of the progenitor of a Type II supernova \citep[e.g.,][]{2009ARA&A..47...63S}.
The secondary star mass is fixed to $M_2=10.8M_\odot$\tm{, corresponding to the initial mass ratio of $0.9$.}
We treat this mass ratio because, as \old{pointed out in} \citet{2002MNRAS.329..897H}, small mass ratios $(q< 0.5)$ cause unstable mass transfer, which makes computations difficult\footnote{It should be noted that {the mass ratio of Galactic O stars follows the uniform distribution,} including binary systems with $q <0.5$ \citep[e.g.,][]{2012Sci...337..444S}.} \citep[see also][]{2017ApJ...840...90O}.
{The initial orbital period is parametrized among }$P_{\rm orb}=1100,1300,1500,1700$ days.
We adopt these values as the period for reproducing the progenitor of a Type II supernova affected by the binary interaction \citep{2017ApJ...840...90O}.
We stop our simulation at the moment of the central carbon depletion $t_{\rm end}=t_{\rm C,dep}$. \old{Here the central carbon depletion is defined as the moment when the mass fraction of carbon in the core falls below $10^{-6}$ (see also Section \ref{subsec:nuclear} \old{and Appendix \ref{app:T_cntr}}).}
This \old{definition} is enough for our purpose because it takes only {$\sim 10$ years} from the moment of the central carbon depletion to the core collapse $t_\mathrm{cc}$ (see Appendix \ref{app:T_cntr} and Figure \ref{fig:T_cntr} for the detail), which is shorter than stellar behaviors under consideration.

In order to construct robust stellar models in terms of spatial resolution, we fix a {\tt MESA} parameter {\tt max\_dq}$=5\times 10^{-4}$. This choice allows us to resolve the mass shell in the star at least $\sim 0.006\,M_\odot$, satisfying the requirement of \citet{2016ApJS..227...22F} that the mass resolution of $\sim 0.01\,M_\odot$ is desirable for the convergence of the evolution of the star during the main-sequence phase. \tm{Hereafter we describe the detailed setups for physical processes directly relevant to our stellar evolutionary simulations. We also refer readers interested in the detailed settings to our inlists uploaded in Zenodo\footnote{\dataset[https://doi:10.5281/zenodo.8385153]{https://doi.org/10.5281/zenodo.8385153}}.}

\subsubsection{Metallicity and opacities}
All the models \old{are} assumed at solar metallicity \citep[$Z=0.02$, where $Z$ is the mass fraction of elements heavier than helium;][]{2009ARA&A..47..481A}.
We \old{employ} the opacity tables from OPAL \citep{1996ApJ...464..943I}.

\subsubsection{Atmosphere}
The radius of a star is defined as the location satisfying $\tau = 2/3$.
\old{In {\tt MESA}, the hydrostatic equilibrium equation is integrated with the Eddington grey $T-\tau$ relation, on the assumption of the plane-parallel limit \citep{1926ics..book.....E}.} This allows us to obtain the surface boundary condition.

\subsubsection{Nuclear reaction network}\label{subsec:nuclear}
We employ the nuclear reaction network provided by MESA under the name {\tt co\_burn.net}, which includes $^{1}$H, $^{3}$He, $^{4}$He, $^{12}$C, $^{14}$N, $^{16}$O, $^{20}$Ne, $^{24}$Mg, and $^{28}$Si.
This is applicable for stellar evolutions up to the oxygen-burning phase \citep{1999ApJS..124..241T}.
We adopt the nuclear reaction rates in the default version of the {\tt MESA}, taken from {\tt NACRE} \citep{1999NuPhA.656....3A} and {\tt JINA REACLIB} \citep{2010ApJS..189..240C}.

We note that after the central carbon depletion, the dynamical timescale of the star can be shorter than the lifetime left, and even the behavior of the primary star as a single star remains a matter of debate \citep[e.g.,][]{2002RvMP...74.1015W, 2011ApJ...733...78A}.
For this reason, we use a simple and computationally inexpensive network sufficient to calculate stellar evolution until the carbon-burning phase.

\subsubsection{Mixing}
For the treatment of convective mixing, we \old{use} the mixing length theory (MLT) approximation \citep{1965ApJ...142..841H} with a mixing-length parameter of $\alpha_\mathrm{MLT} = 2.0$.
The Ledoux criterion for convection is used, and semi-convection following \citet{1985A&A...145..179L} is employed with an efficiency parameter $\alpha_\mathrm{sc} = 1.0$ \citep{2010ApJ...725..940Y}.
We apply the overshooting for the convective core and shell experiencing hydrogen burning, and also for the convective core and shell where no significant burning takes place.
\rs{The overshoot is realized by exponentially decaying the convective diffusion coefficient from the convective boundary according to \citet{2000A&A...360..952H} (see also \citealt{2013ApJS..208....4P} for detail).}
We take an overshooting parameter as $0.016 H_p$, where $H_p$ is the pressure scale height evaluated at the radius near the boundary of the convective core (with $f=0.018$ and $f_0=0.002$).
Although {\tt MESA} has options to include other kinds of mixing processes (e.g., thermohaline mixing, rotational mixing, and extra mixing), these are not considered in this study.

\subsubsection{Wind}
For the stellar wind, we follow the `Dutch' wind scheme implemented in {\tt MESA} with a scaling factor fixed to $1.0$. 
The `Dutch' wind scheme in MESA combines the results from several papers. 
For cool stars with effective temperatures $T_\mathrm{eff} <10^4$ K, we apply the wind scheme in \citet{1988A&AS...72..259D}.
\footnote{\tm{We note that there is still a debate on the uncertainty of the mass-loss rate of cool stars. See e.g., \citet{2020MNRAS.492.5994B} for the recent analysis}.}
This prescription is validated through observations of galactic red supergiants \citep{2011A&A...526A.156M}.
For \old{hot} stars with the effective temperature $T_\mathrm{eff} >10^4$ K, \old{we adopt two formulae depending on the surface mass fraction of hydrogen $X_\mathrm{H,surf}$; we use the formula suggested in \citet{2001A&A...369..574V} for hydrogen-rich stars (\old{$X_\mathrm{H,surf}>0.4$}), 
while we use the recipe of \citet{2000A&A...360..227N} for hydrogen-poor stars (\old{$X_\mathrm{H,surf}<0.4$}).}

\subsubsection{Binary System}
In binary systems, we consider only non-conservative mass transfer.
Our calculations include the loss of angular momentum due to the binary motion, such as mass loss. 
\old{We assume that all of the angular momentum loss in the binary system occurs from the vicinity of the accretor star \citep[see][]{1997A&A...327..620S,2015ApJS..220...15P}.}
This causes the period of the binary system to evolve in time from the initial period, but these detailed dependencies are beyond the scope of this paper.
The Roche lobe radius \old{of the primary star} is calculated following the \citet{1983ApJ...268..368E} method,
\begin{equation}
        \old{R_\mathrm{RL,1}=\cfrac{0.49q^{2/3}}{0.6q^{2/3}+\ln{(1+q^{1/3})}}~a}~,\label{eq:rocheR}
\end{equation}
where \rs{$q= M_1 /M_2$} is the mass ratio and $a$ is the binary separation expressed as
\begin{equation}
        a=\left(\cfrac{G(M_1+M_2)P_{\rm orb}^2}{4\pi^2}\right)^{1/3}~,\label{eq:separation}
\end{equation}
\old{where $G$ is the gravitational constant.} When one of the stars in the binary system initiates Roche lobe overflow, we implicitly compute the mass-transfer rate using the prescription described in \citet{1990A&A...236..385K}.
{Here, we introduce the accretion efficiency parameter $\beta$ \citep{2006csxs.book..623T}, which prescribes the fraction of the gas accreting onto the secondary star relative to the gas lost from the primary star due to the binary stripping.}
In other words, we have the following relationship between the total mass-loss rate of the primary star $\dot{M}_1$, the mass-transfer rate between the binary system $\dot{M}_\mathrm{tr}$, and the mass outflow rate into the CSM $\dot{M}_\mathrm{CSM}$;
\begin{align}
    \dot{M}_1&=\dot{M}_\mathrm{wind,1}+\dot{M}_\mathrm{tr}\label{eq:rate1}~,\\
    \dot{M}_\mathrm{CSM}&=\dot{M}_\mathrm{wind,1}+(1-\beta)\dot{M}_\mathrm{tr}\label{eq:rate2}~,
\end{align}
where $\dot{M}_\mathrm{wind,1}$ is the mass-loss rate due to the stellar wind of the primary star.
We neglect the contribution of the stellar wind from the secondary star. This is because it remains in the main sequence during the simulations so that \tm{the mass and momentum budgets belonging to the stellar wind from the secondary star are smaller than those from the primary star itself and the gas under the transfer.}

\subsection{Hydrodynamics of the CSM formation}\label{subsec:pluto}
We adopt the code {\tt PLUTO} \citep{2007ApJS..170..228M,2012ApJS..198....7M} to solve the equations of hydrodynamics in one-dimensional spherical coordinates. We prepare the simulation domain from the inner boundary radius \old{$r_{\rm in}= V_w (t_{\rm cc} - t_{\rm end}) \simeq 3\times 10^{13}\,{\rm cm}\ (V_w/10\,{\rm km}\,{\rm s}^{-1})$} to $r_{\rm out} = 10\,{\rm pc}$ \old{($\simeq 3\times 10^{19}\,{\rm cm}$)} and divide the domain into 256 meshes in the logarithmic scale where $V_w$ is the value of the velocity of the gas flowing from the binary system. Here we assume $t_{\rm cc} - t_{\rm end} = 10\,{\rm years}$ (see also Section \ref{subsec:mesa}).
As an initial profile we consider the thermodynamic quantities $\rho=1.6\times 10^{-24}\,{\rm g}\,{\rm cm}^{-3}$ \old{(corresponding to the number density of }$\old{n=1\,{\rm cm}^{-3})}, T=10^4\,{\rm K}$ to mimic warm interstellar medium \citep{Draine+2010}. The outflow condition is applied to the outer boundary in the simulation domain.

By injecting the gas from the inner boundary following the model for the mass-loss history obtained in Section \ref{subsec:mesa}, we can construct the hydrodynamical structure of the CSM. We need to assume the value of $V_w$; there is uncertainty as to what the realistic values of the velocity of the gas escaping from the binary system should be.
\old{The reasonable choice would be the velocity comparable to the escape velocity of the central object $(V_w\sim V_\mathrm{esc}\propto R_*^{-1/2}$, where $R_*$ is the stellar radius).}
$V_w\sim\mathcal{O}(10)\,{\rm km}\,{\rm s}^{-1}$ would be expected if the escape velocity of the \old{inflated primary star} gives a main contribution, while $V_w\sim 1000\,{\rm km}\,{\rm s}^{-1}$ might be possible in a case where the gas is coming from the compact secondary star. We examine the value of $V_w = 10, 100,$ and $1000\,{\rm km}\,{\rm s}^{-1}$ and apply these parameters for the simulation of the CSM formation in the binary models in Section \ref{subsec:mesa}.
The CSM density at the inner boundary radius $r_{\rm in}$ is given as
\begin{eqnarray}\label{eq:rho_inj}
    \rho_{\rm inj} = \frac{\dot{M}_1}{4\pi r_{\rm in}^2 V_w}.
\end{eqnarray}
The structure within the inner boundary would be determined by the progenitor activity after \old{$t\sim t_{\rm end}$}, which we do not investigate in detail. 

\old{It is also possible to examine the CSM formation by making use of the analytic treatment on the time-dependent wind dynamics as presented in \cite{2020ApJ...894....2P}. This is applicable as long as the ram pressure of the wind surpasses the thermal pressure of the ambient medium swept by the wind. Once these two pressures become comparable to each other, the ambient medium will suppress the expansion of the wind and the shocked wind gas begins to accumulate at the contact discontinuity between the wind and the medium, deviating from the analytical solution \citep{1977ApJ...218..377W, 2022ApJ...930..143M}. To investigate large-scale structure of the CSM in the framework of binary interaction, we rely on numerical hydrodynamical simulations in this study. A similar method is employed in the context of numerical simulations of supernova remnants \citep{1990MNRAS.244..563T, 1991MNRAS.251..318T, 2005ApJ...630..892D, 2007ApJ...667..226D, 2021ApJ...919L..16Y, 2022ApJ...925..193Y, 2022ApJ...930..143M}. In addition, the choice of the large outer radius of the simulation domain $r_{\rm out} = 10\,{\rm pc}$ allows us to find out the difference between numerical and analytical treatments, which is discussed in Appendix \ref{app:CSM}}.

\section{Result}\label{sec:result}
In this section, we describe the results of our stellar evolutionary simulation and the expected CSM configuration. For convenience, we define the look-back time $t_{\rm lb} = t_{\rm cc} - t$ as an indicator of the evolutionary phase of the star. 

\subsection{Binary evolution}\label{subsec:binary}
\begin{table*}
\begin{center}
    \caption{
Summary of physical quantities in this system, especially for the terminal state of the primary star.
} \label{tbl:result}
  \begin{tabular}{c|cccc} \hline \hline
    initial period & final mass & H-envelope mass & CSM mass for $V_w=10{\rm km}\,{\rm s}^{-1}$ & CSM mass for $V_w=1000{\rm km}\,{\rm s}^{-1}$\\  
    $P_{\rm orb}$ [days] & $M_{1,f}$ [$M_\odot$] & 
    $M_{\mathrm{env},f}$ [$M_\odot$] & 
    $M_\mathrm{CSM}$ [$M_\odot$] & $M_\mathrm{CSM}$ [$M_\odot$]  \\    \hline 
    1100 & 4.8  & 1.3 & 1.4 & 0.3 \\ 
    1300 & 5.9  & 2.4 & 5.0 & 0.6 \\ 
    1500 & 7.2  & 3.7 & 3.8 & 0.9 \\ 
    1700 & 10.8 & 7.3 & 0.05 & 0.02 \\    \hline 
    single star & 10.8 & 7.3 & 0.05 & 0.02 \\
    \hline \hline
  \end{tabular}
\end{center}
\tablecomments{
\tm{The `final mass' refers to the total stellar mass of the primary star and `H-envelope mass' is defined as the amount of the hydrogen envelope in the primary star. The CSM mass is defined as the enclosed mass within the radius at which the CSM distribution reconstructed through the analytical treatment matches well with that through the numerical simulation (see Appendix \ref{app:CSM}). All of the mass budgets are computed with quantities at the end of the simulation (i.e., at the moment of the central carbon depletion of the primary star).}
}
\end{table*}

\begin{figure*}
\centering
  \includegraphics[width=0.49\textwidth]{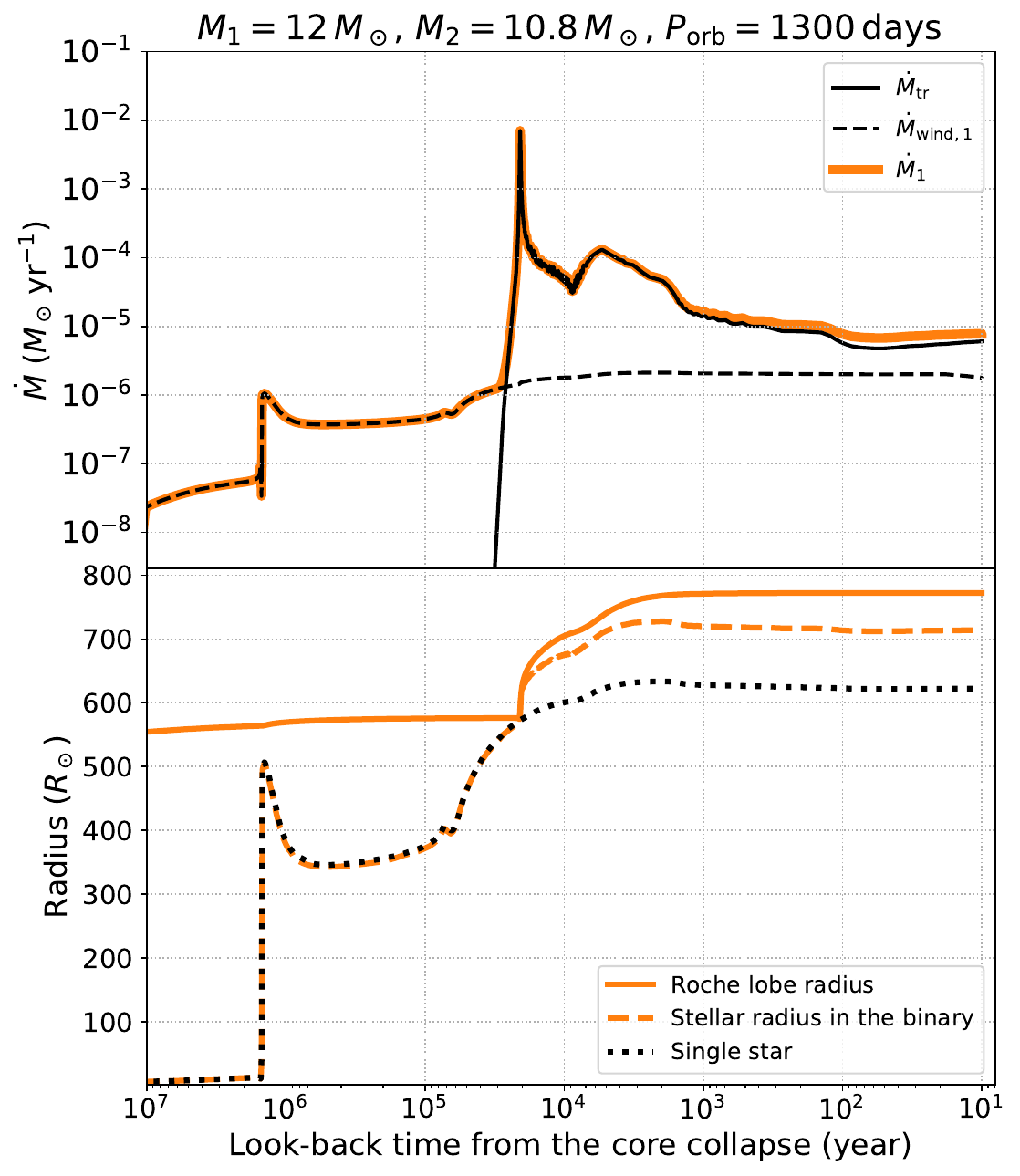}
  \includegraphics[width=0.49\textwidth]{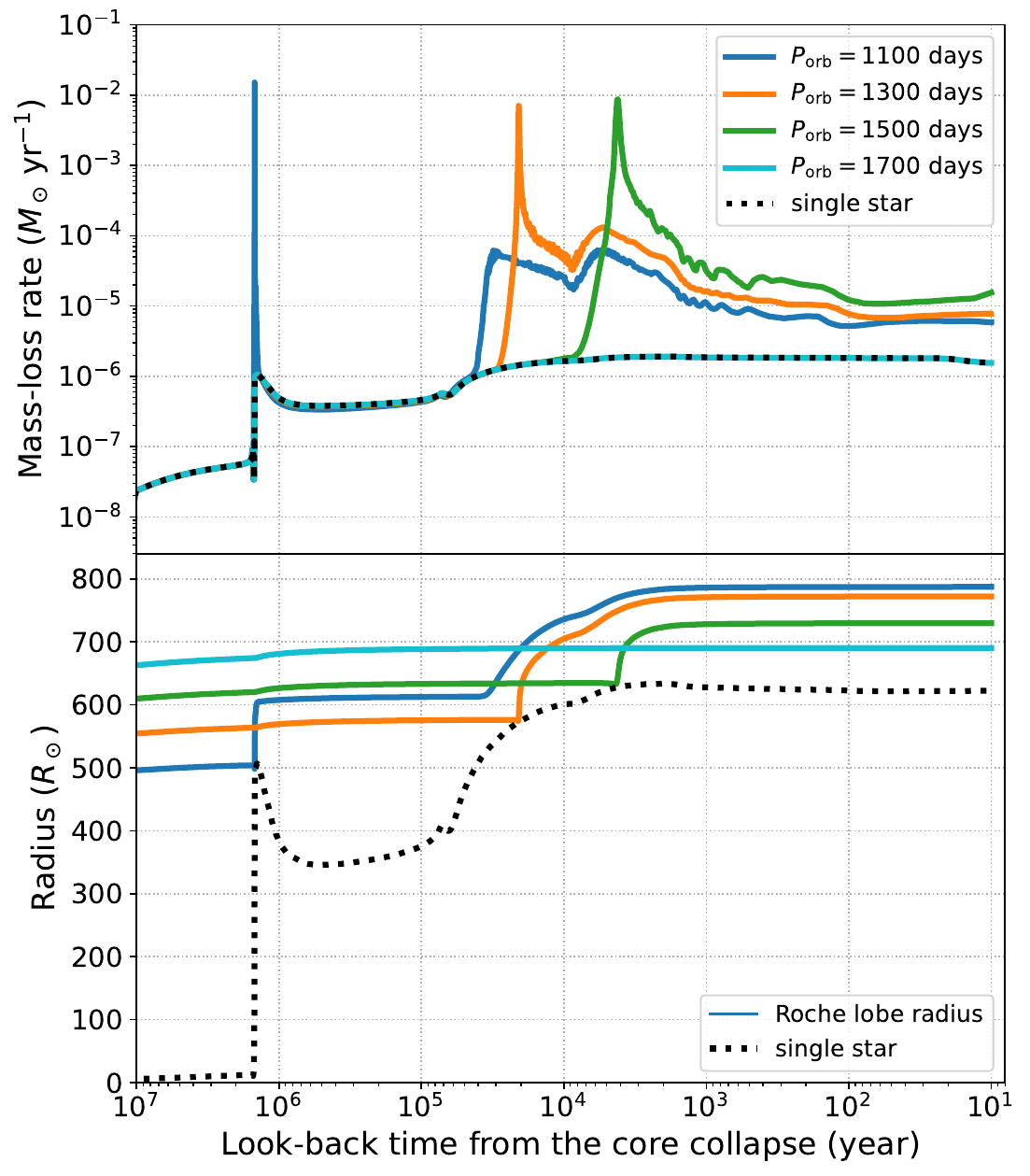}
    \caption{Left: Time evolution of the mass-loss history (top) and the characteristic radii (bottom) as a function of the look-back time for the binary models with $P_{\rm orb}=1300\,{\rm days}$. In the top panel the mass-loss rate is decomposed into the components of the stellar wind and the mass transfer to the secondary star in the top panel. In the bottom panel, the time evolutions of the Roche lobe radius and the stellar radius of the primary star are illustrated, in addition to the $12\,M_\odot$ single star's radius. Right: The total mass-loss rate histories and the Roche lobe radii for the binary models with $P_{\rm orb}=1100, 1300, 1500, 1700\,{\rm days}$ are shown. 
    }
\label{fig:mdot_radius}
\end{figure*}

\tm{Table \ref{tbl:result} shows the summary of the total stellar mass and the hydrogen envelope mass of the primary star, and the CSM mass in our binary models at the end of the simulations. We can observe a tendency for the total mass of the star and the hydrogen envelope to become smaller as the orbital period is shortened, because a tighter binary suffers more strong binary interaction.
We can estimate the ejecta mass of the subsequent SNe at several solar masses for all of the models by subtracting the mass of a newborn neutron star from the final stellar mass.
This could be smaller than the median value of Type II-P SNe containing a large amount of the hydrogen envelope \citep{2022A&A...660A..41M}, but typical for the SNe II got rid of some fractions of the hydrogen envelope, such as Type II-L and II-b SNe \citep{2012ApJ...757...31B, 2016MNRAS.455..423M}.}

Figure \ref{fig:mdot_radius} shows the time evolution of the mass loss rate $(\dot{M}_{1})$ and Roche lobe radius \old{$(R_\mathrm{Rl,1})$} of the primary star in the binary models with the orbital period $P_{\rm orb}=1100, 1300, 1500, 1700\,{\rm days}$. 
The time evolution of the mass-loss rate can be interpreted in the basic framework of the binary evolution \citep[e.g.,][]{2017ApJ...840...90O}. 
We can see that the models with $P_{\rm orb}=1100, 1300, 1500\,{\rm days}$ experience a drastic increase in the mass-loss rate. 
{The timings of the increase in $\dot{M}_1$} coincide with the moment when the primary star fills the Roche lobe radius.
{Our simulations suggest that even without introducing the single star's activity arising eruptions \old{\citep[e.g.,][]{2014ApJ...780...96S, 2017MNRAS.470.1642F,2021ApJ...906....3W,2022ApJ...930..168K}}, an SN progenitor involved in a binary can boost the mass-loss rate from the system at the moment of the expansion of the stellar radius.}

Firstly, we start the detailed discussion in the model with $P_{\rm orb}=1300\,{\rm days}$ as an example. The left panel in Figure \ref{fig:mdot_radius} shows the time evolution of characteristic mass-loss rates and the radii in the primary star. They show that the primary star experiences a mass loss by the stellar wind in the hydrogen-burning phase \old{($t_{\rm lb} \gtrsim 10^6\,{\rm years}$)}.
Then, \tm{due to the increase in the stellar luminosity in addition to the decrease in the effective temperature} after the core hydrogen burning, the stellar wind mass loss rate increases by 30 times around $t_{\rm lb}\sim 10^6\,{\rm years}$, while the Roche lobe overflow does not occur at the time in this model.
At $t_{\rm lb}\sim 2\times 10^4\,{\rm years}$ the helium burning in the stellar core ends, and the primary star expands again (see the bottom right panel in Figure \ref{fig:mdot_radius}).
At this timing, the stellar radius reaches the Roche lobe radius and the binary mass transfer begins. 
The mass transfer rate then \old{rapidly} increases up to $10^{-2}\,M_\odot\,{\rm yr}^{-1}$, and the mass of the primary star reduces down to $M_1\sim 6\,M_\odot$.
The second bump in the mass transfer history can be seen at $t_{\rm lb}\sim 5000\,{\rm years}$.
This could be associated with the single stellar activity of the primary star, since the evolution of the single star with the same initial mass (black dashed line in the left bottom panel of Figure \ref{fig:mdot_radius}) also shows the slight expansion of the stellar radius at $t_{\rm lb}\sim 5000\,{\rm years}$ (see also Section \ref{sec:discussion}).
{After that, the Roche lobe overflow continues until the core collapse, while the mass-transfer rate is regulated down to $\sim10^{-5}\,M_\odot{\rm yr}^{-1}$.}

\old{We note that the primary star in the binary model is inflating more than the single star. This is caused by the mass reduction through the Roche lobe overflow after the full growth of the stellar core. We can understand the behavior of the stellar radius on the basis of the energy balance between the gravitational energy and the internal energy of the star, as explained by the Virial theorem \citep[][\tm{or see Appendix \ref{app:M_R} for the quantitative interpretation}]{1990sse..book.....K}.
The internal energy of the star is mainly determined by the thermodynamical state of the stellar core \citep{2021A&A...656A..58L}. During the Roche lobe overflow in our binary models, the primary star settles in the helium-burning stage, implying sufficient growth of the stellar core. Thus, both the primary star in the binary model and the single star should have similar internal energies at $t_{\rm lb} \lesssim 10^5\,{\rm years}$.
In contrast, the reduction of the stellar mass through mass transfer in the binary results in an increase in the gravitational energy of the star. Consequently, the primary star in the binary can easily expand the radius larger than the star without the companion star.}

\old{Comparing the evolutions of a single star with the star in a binary can be a clue to understanding the critical physical processes in a binary system.} The radius of the primary star in the binary follows the same path as that evolving as a single star until it first fills the Roche lobe radius.
This implies that the timing when a star first fills the Roche lobe radius can be roughly understood through the comparison between the Roche lobe radius and the radial evolution in the single stellar evolution \old{(see later)}.

The right panel of Figure \ref{fig:mdot_radius} shows the dependence of mass loss properties on the orbital period $P_{\rm orb}$.
In all binary models, the primary star will eventually become a Type II SN progenitor because the primary star retains the hydrogen envelope until the core collapse (see Table \ref{tbl:result}).
We can see the trend that the longer the initial orbital period $P_{\rm orb}$, the further the timing of the Roche lobe overflow is delayed.
When the initial orbital period is too long ($P_{\rm orb}>1700\,{\rm days}$), the mass loss rate is roughly the same as the mass loss rate in single stellar evolution, since the binary does not experience Roche lobe overflow (see dashed black line and solid cyan line in the right panel of Figure \ref{fig:mdot_radius}).

{It is possible to compare} the binary models with $P_{\rm orb}=1100,\,1300,\,{\rm and}\,1500\,{\rm days}$ to the single stellar evolution.
All binary models follow the same evolutionary path as the single stellar model until $t_{\rm lb}\sim 10^6\,{\rm years}$.
The model of $P_{\rm orb}=1500\,{\rm days}$ follows almost similar evolution to the model of $P_{\rm orb}=1300\,{\rm days}$ discussed above, except that the Roche lobe overflow starts slightly later due to the larger Roche lobe radius.
The model with $P_{\rm orb}=1100\,{\rm days}$ undergoes a Roche lobe overflow after the end of the core hydrogen burning at $t_{\rm lb}\sim 10^6\,{\rm years}$, which leads to the release of about $\sim 7\,M_\odot$ of the hydrogen outer layer.
It then experiences another Roche lobe overflow at $t_{\rm lb}\sim \, 3\times10^4\,{\rm years}$, following a path different from that in single stellar evolution.
However, we should emphasize that it is no coincidence that
$P_{\rm orb}=1100\,{\rm days}$ model exhibits the mass-loss enhancement at $t_{\rm lb}\sim10^4\,{\rm years}$ similarly to the other binary models ($P_{\rm orb}=1300, 1500\,{\rm days}$). 
Normally, helium burning in the stellar core ends at $\sim10^4$ years before the core collapses \citep[e.g.,][]{2002RvMP...74.1015W}. 
At this time, the primary star expands again, and thus we can see an enhancement of mass loss.

We note the model of $P_{\rm orb}=1700\,{\rm days}$ follows 
the same mass loss history as the single stellar model 
\rs{since the primary star radius remains sufficiently smaller than its Roche lobe radius in the whole of the lifetime of the star.}
Therefore, while we conducted the CSM reconstruction even for the binary model with $P_{\rm orb}=1700\,{\rm days}$, we do not discuss the result of the model in the next section.

\subsection{Reconstruction of CSM}\label{subsec:csm}

\begin{figure}
\centering
  \includegraphics[width=0.49\textwidth]{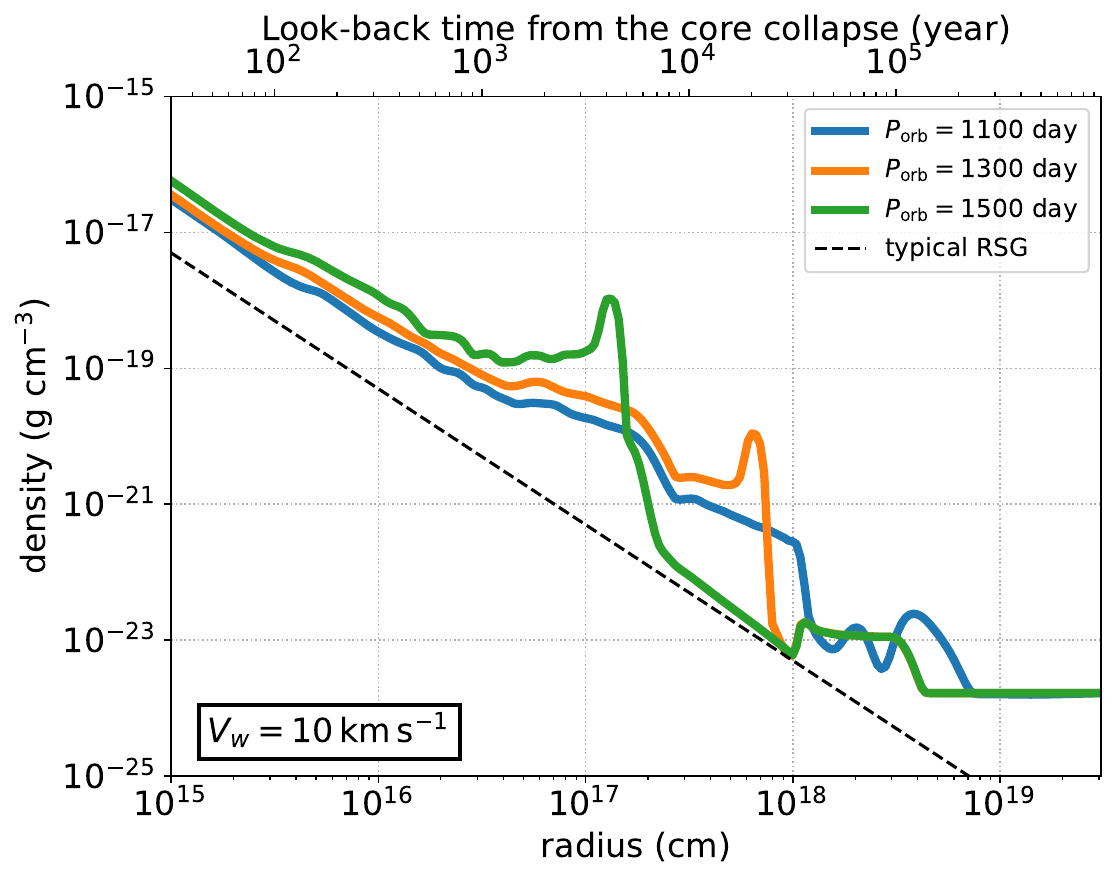}
    \caption{Density structures of the CSM in the binary models with $P_{\rm orb}=1100, 1300, 1500$ days reconstructed on the assumption of the wind velocity $V_w = 10\,{\rm km}\,{\rm s}^{-1}$. The dashed line stands for the typical steady wind distribution expected for a red supergiant ($\dot{M}=10^{-6}\,M_\odot{\rm yr}^{-1}$, $V_w=10\,{\rm km}\,{\rm s}^{-1}$).
    }
\label{fig:Porb_dependence}
\end{figure}

Figure \ref{fig:Porb_dependence} shows the density structures of the CSM for the binary models assuming $V_w=10\,{\rm km}\,{\rm s}^{-1}$. We can see that the shell-like distribution is standing out at $\sim \mathcal{O}(10^{17})\,{\rm cm}$ in the models with $P_{\rm orb}=1300, 1500\,{\rm days}$. The corresponding look-back time is around $t_{\rm lb} \sim 10^4\,{\rm years}$ and it is easily found that at that time the mass-loss rate is enhanced by orders of magnitude, compared to that expected for the steady wind of a red supergiant (RSG). \old{As for the model with $P_{\rm orb}=1100\,{\rm days}$, not shell-like but the cliff-like structures in the CSM are appearing at the radii of $2\times 10^{17}\,{\rm cm}$ and $10^{18}\,{\rm cm}$. These features are associated with the mass-loss enhancements at $t_{\rm lb}\sim 5000\,{\rm years}$ and $3\times 10^4\,{\rm years}$, respectively.} We suggest that binary interaction can give rise to the formation of the shell-like \old{and cliff-like} CSM structures.

\rs{We confirm that there is no mixing of heavy elements into the hydrogen envelope of the primary stars within the parameter range of this study.
Also as can be seen from Table \ref{tbl:result}, the mass release from the primary star happens only from the hydrogen layer, not from the helium core.
Therefore, the CSM composition is expected to be similar to that of Sun.}

\begin{figure}
\centering
  \includegraphics[width=0.49\textwidth]{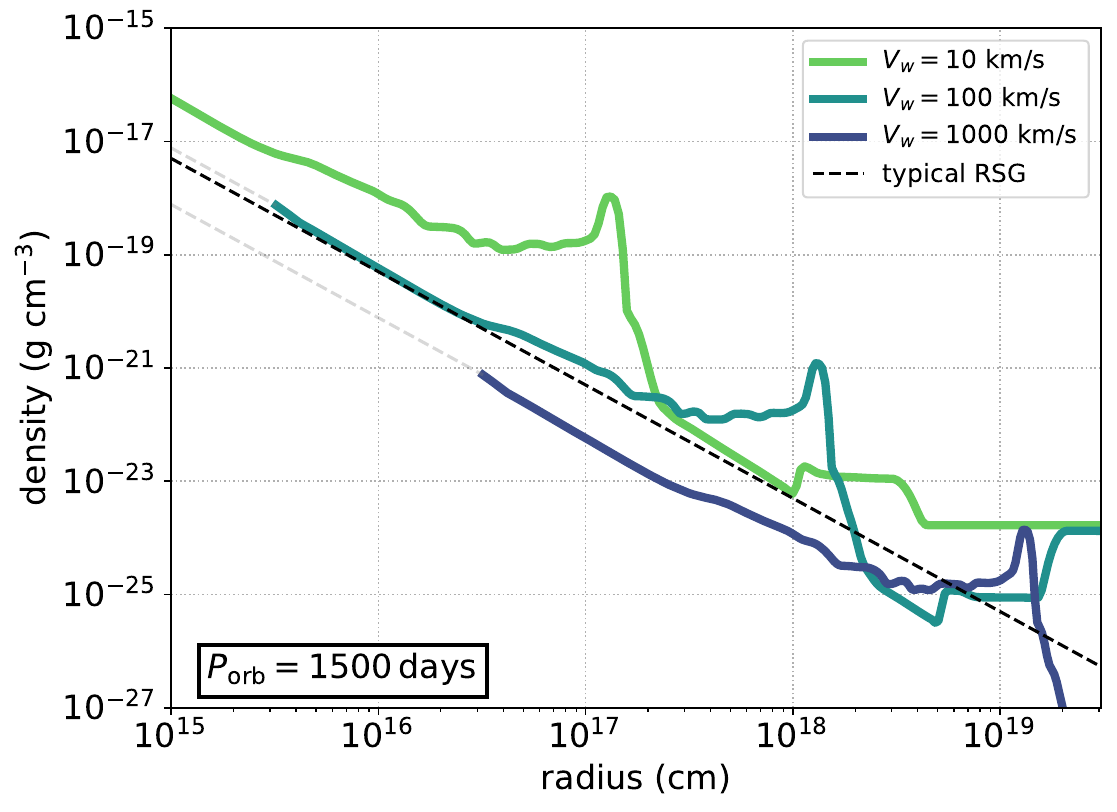}
    \caption{Density structures of the CSM in the binary models of $P_{\rm orb}=1500$ days with the velocity varied among $10, 100,$ and $1000\,{\rm km}\,{\rm s}^{-1}$. \old{The dashed line within $r<3\times 10^{15}\,{\rm cm}$ in the model of $V_w=100\,{\rm km}\,{\rm s}^{-1}$ and $r<3\times 10^{16}\,{\rm cm}$ in the model of $V_w=1000\,{\rm km}\,{\rm s}^{-1}$ denote the components of the CSM with their corresponding look-back timescale less than $t_{\rm cc}$, during which we do not explore.} The black dashed line shows a steady wind structure expected for a red supergiant, the same as the one in Figure \ref{fig:Porb_dependence}.}
\label{fig:Vwind_dependence}
\end{figure}

The dependence of the CSM structure on the wind velocity is also worth investigating, which is shown in Figure \ref{fig:Vwind_dependence}. We can see that higher wind velocity leads to the widely distributed and thin-density structure of the CSM. As $V_w$ becomes faster, the wind can reach the lengthscale farther from the SN progenitor, while its density becomes smaller (see equation \ref{eq:rho_inj}). We note that the shell-like component in the model with slower wind velocity lies in the smaller lengthscale.

\old{The variety of hydrodynamical structures of CSM would give rise to observational characteristics of subsequent SNe, such as the excess in the optical emission and the re-brightening of radio emission in the late phase of SN-CSM interaction \citep[e.g.,][]{2020ApJ...900...11W,2022ApJ...936..111K,2023ApJ...945L...3M}.}
Here we discuss the possibility that our results can give explanations for \old{characteristics of SNe indicated by late-phase observations.} If the wind velocity is as slow as orders of $\sim 10\,{\rm km}\,{\rm s}^{-1}$, then the location of the mass shell observed in the model with $P_{\rm orb}=1300, 1500\,{\rm days}$ falls in $\sim 10^{17}\,{\rm cm}$. This could be an origin for the observational signature of late-time SN-CSM interaction seen in several Type II SNe because the SN shock begins interacting with the shell at the radius of $\sim 10^{17}\,{\rm cm}$ a few years after the explosion.
\tm{To quantitatively discuss the variation of the CSM density in the radial direction, mass-loss rate of SN progenitor can serve as an indicator of the density variation of the CSM. \cite{2021ApJ...918...34M} argues that the variation in the mass-loss rate only by a factor can emerge as an observational peculiarity in radio light curves of SNe. \citet{2023ApJ...945L...3M} also suggests that the elevation of the mass-loss rate by an order of magnitude could be an explanation for the radio re-brightening of SNe. Other cases are that \citet{2020ApJ...900...11W} and \citet{2022ApJ...936..111K} ascribe the origin of the excess in the optical light curves as the existence of the CSM with their mass-loss rates not so much higher ($\dot{M}\sim 10^{-7}-10^{-6}\,M_\odot{\rm yr}^{-1}$), but these densities are enough for the appearance of the observable CSM interaction features. As for our binary models reproducing shell-like CSM structures, their density variations range from factors to an order of magnitude, and we expect these variations enough to be ascribed as the origin for the CSM interaction features observed in late-phase SNe.}

The cliff-like structure seen in the model of $P_{\rm orb}=1100\,{\rm days}$ can be also intriguing since the cliff-like structure may lead to the attenuation of the CSM interaction signature in the late phase of SNe. Indeed, \citet{2007ApJ...671.1959W} reported a steep dimming in the radio emission from SN 1993J $\sim 10$ years after the explosion.
\tm{They further suggested that the mass-loss rate increased up to at least 3 times $\sim 8000$ years before the explosion, corresponding to the lengthscale of the CSM of $2\times10^{17}\,{\rm cm}$. Our model with $P_{\rm orb}=1100\,{\rm days}$ exhibits a density variation comparable to or even higher than the above implications, supporting that our binary models can be likely to give rise to CSM interaction features sufficiently prominent to be traced by observations.}
We also note that the binary interaction scenario is consistent with the fact that SN 1993J is a Type IIb SN originating from a massive star probably suffered from the stripping of the envelope by a companion star \citep[e.g.,][]{2004Natur.427..129M}.

\old{
In addition to implications from SN observations, we also note that direct observations of late-type stars in our galaxy indicated the existence of the multiple-shell structure in the circumstellar environment due to the binarity \citep{1997A&A...326..300M, 1999A&A...349..203M}. This supports that binary interaction can be a feasible origin of the emergence of the shell-like structure in the circumstellar environment of stars \citet{1997A&A...326..300M, 1999A&A...349..203M}.
}

\section{Discussion}\label{sec:discussion}
\old{In the previous section, we have mainly discussed the relationship between stellar evolution in a binary and the CSM formation. Yet, there are some physical processes and parameter spaces that have not been explored or considered in detail.
In this section, we discuss the mass dependence of the primary star (Section \ref{subsec:diversity}), the binary evolution with the short orbital period reproducing stripped-envelope SNe (Section \ref{subsec:SESNe}), and some caveats relevant to CSM formation associated with binary interaction (Section \ref{subsec:other}).}

\subsection{Diversity of the binary evolution and its dependence on the mass of the primary star}\label{subsec:diversity}
In this study, we fixed the value of the initial primary mass to clarify the physics of the CSM formation.
Our results show that the CSM properties depend on the evolution of the primary star radius $R_1$, as well as several physical parameters such as $P_\mathrm{orb}$ and $V_w$.
Obviously, the radial evolution of the primary star depends on the choice of the initial primary mass, and the variation of the primary mass expands the diversity of stellar evolution itself.
{Here, we present a series of models in which the initial primary mass is parametrized, and discuss the effect of the variation of the primary mass on the behavior of the mass-loss history and the resultant CSM structure qualitatively.}

\rs{Here we note the purpose of this section. We will show how different primary masses, i.e., different stellar evolutions of the primary stars, give rise to different mass loss activities and CSM structures. 
This is because the evolution of mass loss due to binary interaction strongly reflects the stellar evolution of the primary star peculiar to itself, as studied for the case of the primary star mass of $M_1 = 12M_\odot$ in Section \ref{sec:result}.
Detailed analysis of the results in this section would necessitate individual stellar evolution calculations, which are beyond the scope of this paper. 
Hence, we just limit our discussion to showcasing an example of the diversity resulting from different primary star masses.}

\begin{figure}
\centering
  \includegraphics[width=0.49\textwidth]{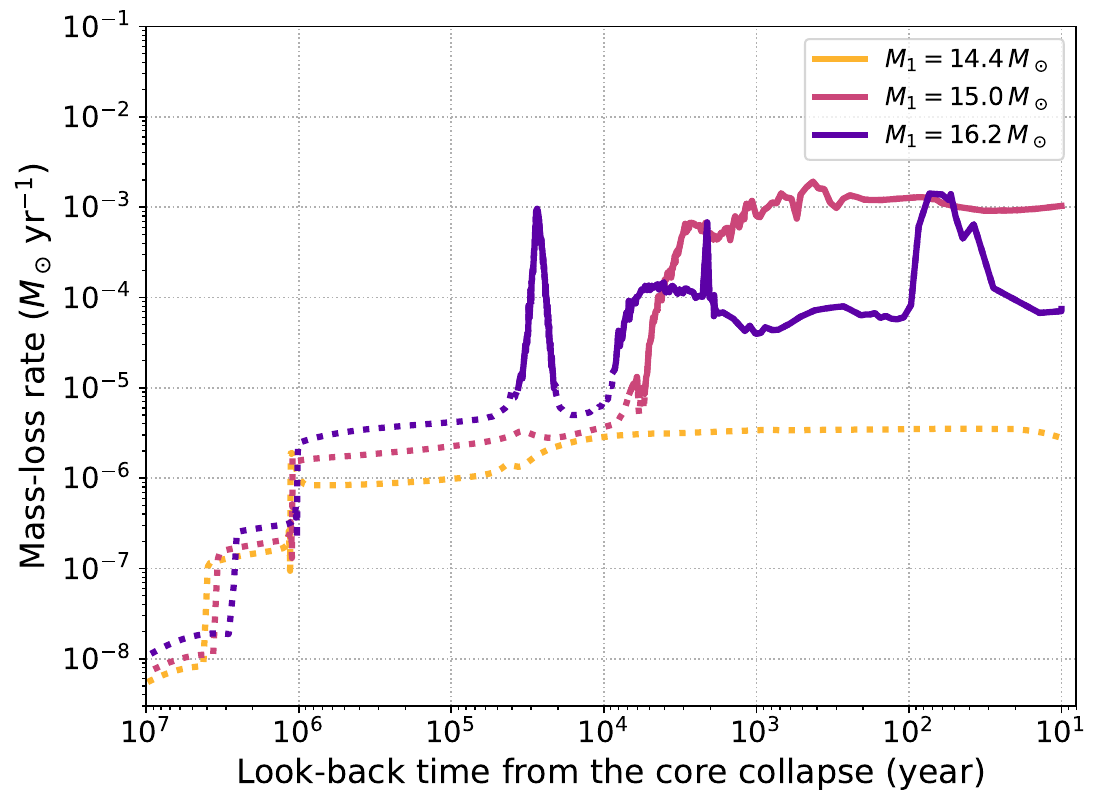}
    \caption{Time evolutions of mass-loss rates in binary systems with $M_1=14.4\,M_\odot, 15.0\,M_\odot$, and $16.2\,M_\odot$ \old{are plotted by dotted lines. In addition, the epochs are highlighted by solid lines when the mass-transfer rate is greater than the mass-loss rate of the stellar wind.} The secondary star's mass and the orbital period are set to $ M_2=13.5\,M_\odot$ and $P_{\rm orb}=1900\,{\rm days}$, respectively.}
\label{fig:confinedMdot}
\end{figure}

Figure \ref{fig:confinedMdot} shows the time evolution of the mass loss rate $(\dot{M}_{1})$ adopted for the primary star masses $M_1=14.4,\,15,\,16.2\,M_\odot$. 
Other physical/binary parameters are fixed; the orbital period $P_{\rm orb}=1900\,{\rm days}$ and the secondary mass $M_2=13.5\,M_\odot$. 
We have confirmed that in all of these models, the primary star will eventually become a Type II SN progenitor encompassed by a hydrogen envelope.
We can see that the models with $M_1=15\,M_\odot$ experience the drastic increase in the mass-loss rate at $t_{\rm lb}\sim \mathrm{a~few}\times10^3\,{\rm years}$.
Also, the models with $M_1=16.2\,M_\odot$ shows the multiple episodes in the mass-loss history around $t_{\rm lb}\sim 3\times10^4$, $10^4$ and $10^2\,{\rm years}$.
On the other hand, in the models with $M_1=14.4\,M_\odot$, no mass-loss gain is observed even though the same orbital period and secondary mass are employed.

\old{We suppose that the differences among the three models are attributed to the behaviors of the primary star.} In the evolutionary stage after the carbon burning, the stellar radius \old{can be variable} due to single stellar activity, such as shell flashes and core convection. When the star is evolving as a single star, these activities would not be reflected in the mass-loss history unless they induce mass loss from the surface of the star \citep[e.g.,][]{2012MNRAS.423L..92Q, 2014ApJ...780...96S}.
This is true even when \old{the primary star is in a binary but its Roche lobe radius is sufficiently larger than the stellar radius (the case of $M_1=14.4\,M_\odot$).}
However, when the Roche lobe radius becomes comparable to the stellar radius of the primary star, the radial variation of the star would be reflected in the mass loss due to Roche lobe overflow, and thus the complicated evolution of the mass-loss rate would be realized.
\old{In other words, binary interaction plays a role in enhancing the stellar behavior of the primary star as a form of mass-transfer history. We expect that the difference in the evolution of the stellar radius between $M_1=15\,M_\odot$ and $16.2M_\odot$ results in the difference in the mass-transfer history.}
We propose that single stellar activities that involve the expansion of the stellar radius but do not induce mass eruptions, may also contribute to mass loss if the star is in a binary system. 
Thus, understanding the massive star's evolution after the carbon-burning phase would be important even for seeking the origin of the dense-CSM structure around SN progenitors.

The mass-loss histories drawn in Figure \ref{fig:confinedMdot} are informative even for the formation of the resultant CSM structure. As examples, we take up the models with $M_1=15.0\,M_\odot$ and $M_1=16.2\,M_\odot$. First, \old{we propose that binary interaction can be one of the origins producing dense CSM inferred in Type IIn SNe.} We can see that the mass-loss rate seen in the model with $M_1=15.0\,M_\odot$ is elevated to $\sim 10^{-3}\,M_\odot\,{\rm yr}^{-1}$ from a few thousand years before the explosion. This is comparable to those previously inferred for Type IIn SNe \citep[e.g.,][]{2012ApJ...744...10K, 2014MNRAS.439.2917M}, and in good agreement with the discussion in \cite{2017ApJ...840...90O}. If we consider the slow velocity of the CSM ($V_w=10\,{\rm km}\,{\rm s}^{-1}$), the CSM in this model is expected to have a cliff-like structure; a steep density drop at the radius of $\sim 10^{17}\,{\rm cm}$. This lengthscale may be consistent within a factor with the implication of \cite{2016ApJ...832..194K}, where some Type IIn SNe are suggested to be encompassed by a torus-like CSM truncated at around several times of $10^{16}\,{\rm cm}$.

\begin{figure}
\centering
  \includegraphics[width=0.49\textwidth]{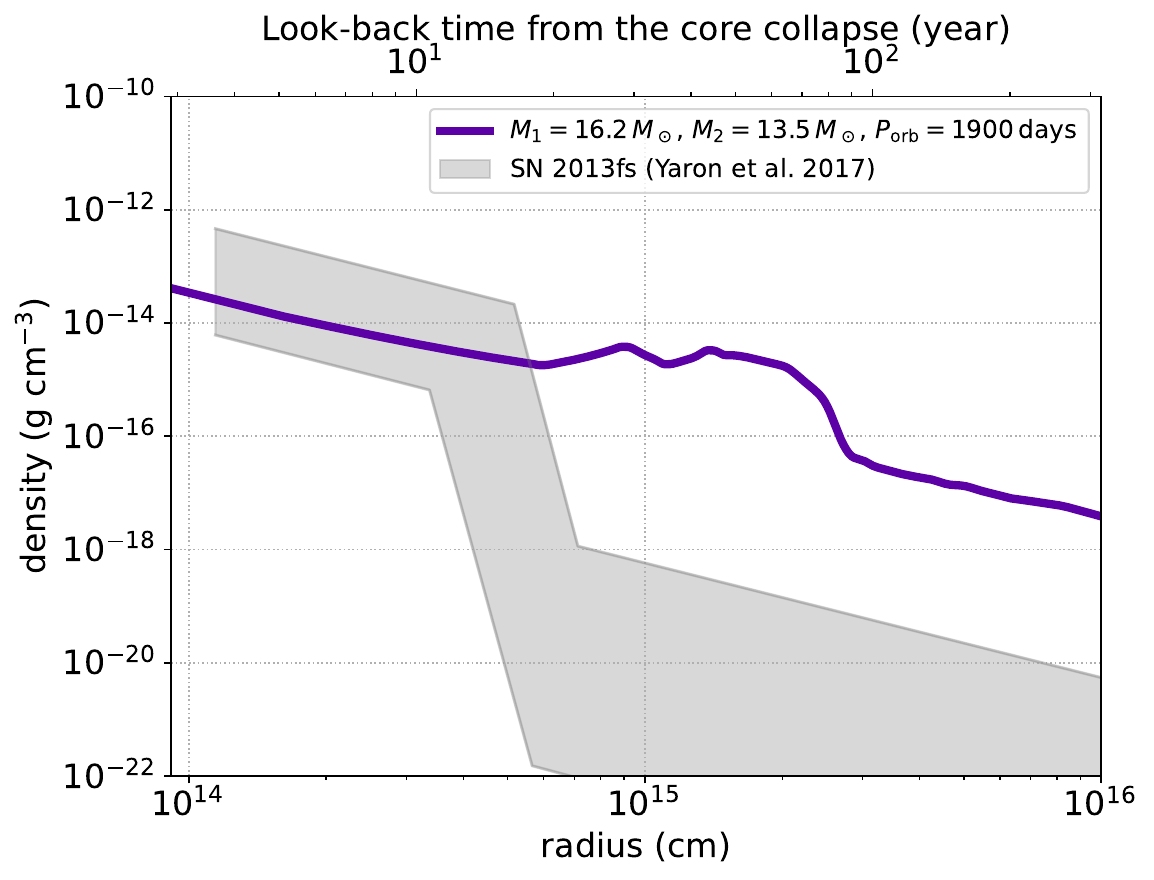}
    \caption{The synthesized CSM density structure for the model with $M_1=16.2\,M_\odot, M_2=13.5\,M_\odot$, and $P_{\rm orb}=1900\,{\rm days}$, assuming the wind velocity $V_w=10\,{\rm km}\,{\rm s}^{-1}$. Gray shaded region illustrates the possible parameter space for the CSM properties inferred for SN 2013fs \citep{2017NatPh..13..510Y}.}
\label{fig:confinedCSM}
\end{figure}

\old{Another suggestion is that binary interaction in the final evolutionary phase can be a possible explanation of the signature of the CSM interaction observed in infant SNe.
Figure \ref{fig:confinedCSM} shows the synthesized CSM structure for the model with $M_1=16.2\,M_\odot$.}
The mass-loss history of this model is characterized by multiple enhancements of the mass-loss rate. 
The last mass-loss episode happens at $t_{\rm lb} \sim 50\,{\rm years}$, and the gas expelling from the binary system at that episode would reach out to $\sim 10^{15}\,{\rm cm}$. We remark that this lengthscale \old{may be} compatible with that inferred for the confined CSM seen in some infant Type II SNe \old{(see e.g., \citealt{2017NatPh..13..510Y} for SN 2013fs, \citealt{2022ApJ...924...15J} for SN 2020tlf, and \citealt{2023arXiv231010727Z} for SN 2023ixf).}
We note a caveat that our model is not quantitatively tuned to the observational implication \old{represented by} SN 2013fs, and the discussion should be limited qualitatively.
Nevertheless, we advocate that the binary interaction episode arising $\lesssim 100\,{\rm years}$ before the explosion can leave an inhomogeneous structure in the vicinity of the SN progenitor \old{and it may contribute to the unique observational features of infant Type II SNe such as flash spectroscopy and early excess of optical light curve \citep{2017NatPh..13..510Y, 2017ApJ...838...28M, 2017MNRAS.469L.108M, 2022ApJ...924...15J}.}

\subsection{\old{Reproducing mass-loss histories of stripped-envelope SN progenitors}}\label{subsec:SESNe}
\old{A stripped-envelope SN is indicated to originate from a binary system \citep[e.g.,][]{2016MNRAS.457..328L,2019NatAs...3..434F}.
There are also suggestions that some stripped-envelope SNe produce a sign of the interaction between the massive shell-like CSM in the late phase of the SNe \citep{2017ApJ...835..140M, 2018MNRAS.478.5050M, 2021ApJ...923...32B}, and this implication looks consistent with our results.
In fact, some binary evolutionary simulations with short orbital periods of $P_{\rm orb}\lesssim 10\,{\rm days}$ through Case A or Case BB mass transfers have been successful in reproducing models for stripped-envelope SN progenitors \citep{2017ApJ...840...90O,2019MNRAS.486.4451G,2020A&A...637A...6L}, although there has been less attention to mass-transfer histories.}

\old{We attempted binary evolutionary simulations of such a short-period system with the same stellar masses of the stars by using the mass-transfer prescription following \citet{2002MNRAS.329..897H}. We found that the mass-transfer rate is intensely fluctuating by orders of magnitude within short timescales. This fluctuation has been observed in the previous works \citep[see Figure 8 in][]{2015ApJS..220...15P} and it prevents us from understanding the features of the mass-transfer history in a tight binary, which we will postpone in future work. However, it may be notable that \citet{2022ApJ...940L..27W} conducted the stellar evolutionary simulations of low-mass He stars accompanied by a neutron star, and the obtained mass-loss rates therein are not fluctuated.}

\subsection{\old{Other physical processes affecting CSM formation}}\label{subsec:other}
\old{This study aims to construct CSM models originating from binary interaction, but there are several assumptions and other possible physical processes that could affect our result. This section discusses their roles in characterizing CSM properties and assesses their importance.
\begin{itemize}
\item {\bf The timing of the termination of stellar evolutionary simulations\\}
We assume that the primary star experiences core collapse $t_{\rm end} - t{\rm cc}=10\,{\rm years}$ after entering the central carbon depletion stage. Notice that our definition of the central carbon depletion stage is the phase when the mass fraction of carbon in the core falls below $10^{-6}$. Based on this definition, we believe that the time span from the central carbon depletion stage to the core collapse would not be largely modified depending on stellar quantities by orders of magnitude. Actually, our definition of the central carbon depletion stage corresponds to the epoch just before the ignition of oxygen burning in the core \citep{2010ApJ...718..357T}. Even if there are variations in $t_{\rm end} - t{\rm cc}$ by factors, it should affect the neighborhood circumstellar environment of SN progenitors with the lengthscale of $V_w(t_{\rm end} - t{\rm cc}) \sim 3\times 10^{14}\,{\rm cm}$ on the assumption of $V_w=10\,{\rm km}\,{\rm s}^{-1}$. This is sufficiently shorter than the characteristic CSM lengthscales discussed in Section \ref{subsec:csm}.
\item {\bf Thermodynamical state of external interstellar medium\\}
This study assumes warm interstellar medium ($\rho=1.6\times 10^{-24}\,{\rm g}\,{\rm cm}^{-3}, T=10^4\,{\rm K}$) as an initial condition of simulations of CSM formation, whereas it is known that interstellar medium can take various phases including dense or dilute environment \citep{Draine+2010}. However, if we focus on observational timescale of SNe and lengthscale of CSM ($t\lesssim\, {\rm a\ few\ years\ or\ }r\lesssim 10^{17}\,{\rm cm}$), it does not matter because the characteristics of CSM is predominantly determined by the bulk properties of the outflow from the Roche lobe overflow. If anything, the difference of the interstellar medium state affects the large-scale structure of the CSM that may influence the evolution of supernova remnants \citep{2021ApJ...919L..16Y, 2022ApJ...925..193Y, 2022ApJ...930..143M}. \tm{We also investigate CSM formation properties with the different initial conditions such as cold ($\rho=1.6\times 10^{-22}\,{\rm g}\,{\rm cm}^{-3}, T=10^2\,{\rm K}$) and hot ($\rho=1.6\times 10^{-26}\,{\rm g}\,{\rm cm}^{-3}, T=10^6\,{\rm K}$) interstellar medium. We confirm that the choice of the thermodynamical state of the interstellar medium does not have an influence on the resultant CSM configuration in the region of which we are interested ($r\lesssim 10^{18}\,{\rm cm}$).}
Basically, the existence of the interstellar medium affects the location of the contact discontinuity with the CSM, which is determined by the balance between the thermal pressure of the shocked interstellar medium and the ram pressure of the wind \citep[see e.g.,][]{1977ApJ...218..377W, 2022ApJ...930..143M}. When we consider cool and dense interstellar medium, the contact discontinuity should be located in the inner region due to the high thermal pressure of the shocked interstellar medium. If the interstellar medium is hot and more dilute, then the CSM would extend farther in the outer region.
\end{itemize}
}

\old{
\begin{itemize}
\item {\bf Asphericity of the outflow:\\}
In this study, the outflows (stellar wind and the outflow from the Roche lobe) are assumed to be spherically symmetric, but non-spherical structures of the CSM can be realized.
However, some studies suggested that if the outflow is via the Lagrangian point of the binary system, the CSM may be distributed along the equatorial plane with the spiral structure \citep{1975ApJ...198..383L, 1979ApJ...229..223S, 2016MNRAS.455.4351P}.
Discussing the ejection mechanism and the resultant morphology of the outflow from the Roche lobe overflow would require multi-dimensional hydrodynamical models, which is beyond the scope of this paper.
Nevertheless, we note that if the aspherical structure of the CSM is realized in SNe, they may leave unique signatures on radiative properties of SNe \citep[e.g.,][]{2019ApJ...887..249S}. 
We also mention that the circumstellar environment around red supergiants can be variable and even clumpy \citep[e.g., VY Canis Majoris;][]{2009AJ....137.3558S}.
This implies that if the stellar wind from the primary has dense clumpy components, it can affect the multi-dimensional configuration of the CSM. 
\end{itemize}
}

\section{Summary}\label{sec:summary}
In this paper, we have calculated mass-loss rates from a massive stars' binary system on the basis of non-conservative mass transfer during Roche lobe overflow. 
We demonstrate that the mass-transfer rate can vary with time according to the expansion or contraction of the stellar radius.
If the Roche lobe overflow begins immediately after the exhaustion of the nuclear burning fuel in the core (e.g., hydrogen or helium), the mass-transfer rate can be intensively enhanced. In the framework of the non-conservative mass transfer in the binary, the time variability of the mass-transfer rate would be directly associated with the spatial inhomogeneity of the CSM density structure. By making use of hydrodynamics simulations we also showed that the abovementioned enhancement of the mass-transfer rate can emerge as the shell-like or cliff-like structures in the CSM. Particularly if the wind velocity is as slow as orders of $10\,{\rm km}\,{\rm s}^{-1}$, {the characteristic radii in these CSM structures fall on $\sim 10^{17}\,{\rm cm}$ and could contribute to the observational signatures traced by long-term observations of SNe.}

\old{Furthermore, we discussed the plausible processes that may yield the diversity of the CSM around SN progenitor, such as the choice of the primary mass.} We found the stellar parameter set that can reproduce the significant increase in the mass-transfer rate, which could be \old{comparable with the values implied} for Type IIn SNe. Another parameter set is found that reproduces the mass-loss episode in the last $100$ years before the explosion. This allows us to synthesize the inhomogeneous structure with the length and density scales compatible with the confined CSM proposed for infant Type II SNe, although fine tuning of the binary parameters would be required.
\old{We also mentioned other possible processes that can have influence on the configuration of the CSM.}
Our binary evolution models would highlight the possible scenario that explains the diversity of the CSM morphology inferred for SN progenitors.

\software{
{\tt MESA} revision {\tt 15140} \citep{2011ApJS..192....3P,2013ApJS..208....4P,2015ApJS..220...15P,2018ApJS..234...34P,2019ApJS..243...10P}, 
{\tt PLUTO} \citep{2007ApJS..170..228M,2012ApJS..198....7M}.
\old{Inlists used to create the models are available on Zenodo under an open-source Creative Commons Attribution license: 
\dataset[https://doi:10.5281/zenodo.8385153]{https://doi.org/10.5281/zenodo.8385153}.}}

\section*{Acknowledgments}
The authors thank Ryoma Ouchi and Keiichi Maeda for providing the original calculation code, and Yudai Suwa, Kenta Hotokezaka, and Daisuke Toyouchi, \old{Ke-Jung Chen, and Nicolas Mauron} for their helpful comments and suggestions.
\old{The authors are also grateful to the anonymous referee for his/her comments improving our manuscript.}
This work has been supported by the Japan Society for the Promotion of Science (JSPS) KAKENHI grants 20H01904 (TM), 21K13964, and 22KJ0528 (RS).
\old{This research is supported by the National Science and Technology Council, Taiwan under grant No. MOST 110-2112-M-001-068-MY3 and the Academia Sinica, Taiwan under a career development award under grant No. AS-CDA-111-M04.}
Numerical simulations in this study were carried out in the supercomputer cluster Yukawa-21 equipped with Yukawa Institute for Theoretical Physics.

\restartappendixnumbering
\appendix
\section{lifetime of the star \old{at the carbon depletion stage}}\label{app:T_cntr}
To simplify the {nucleosynthesis} network calculations, we have approximated the {carbon depletion time} to the time when the primary star undergoes the core collapse (see Section \ref{subsec:mesa}). To check the validity of this assumption, we calculate the further evolution of a star with $M_{\rm ZAMS}=12\,M_\odot$ after the carbon depletion in the core.

\begin{figure}
\centering
  \includegraphics[width=0.49\textwidth]{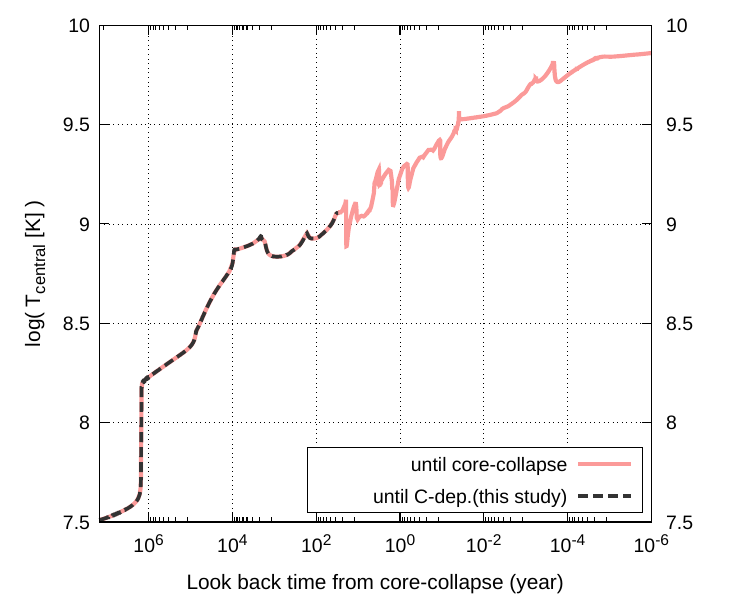}
    \caption{Time evolution of the central temperature of a single star with $M_{\rm ZAMS}=12\,M_\odot$ as a function of look-back time from the core collapse. The red line shows the evolution until the core collapse of the Fe core, while the black dashed line indicates the same but until the carbon depletion, to highlight the treatment of our simulation.}
\label{fig:T_cntr}
\end{figure}

Figure \ref{fig:T_cntr} shows the result of the calculation up to the central carbon depletion stage adopted in this study (dashed black line) and up to the time of the core collapse (solid red line).
\old{The horizontal axis shows the logarithmic scale of the look-back time of the star measured from the core collapse.}
The `central carbon depletion stage' is defined as when the central carbon abundance depletes below $10^{-6}$, as described in the text, and `core-collapse' is defined as when the infall-velocity in the central iron core exceeds $10^7\,{\rm cm}\,{\rm s}^{-1}$. 

As can be seen in Figure \ref{fig:T_cntr}, the result up to the central carbon depletion stage is stopped at a central temperature of $\log T_\mathrm{c}=9.1~(\mathrm{i.e.~}T\approx1\times10^9~\mathrm{[K]})$, while the result up to the core collapse stage reaches the central temperature of $\log T_\mathrm{c}=9.8~(\mathrm{i.e.~}T_\mathrm{c}>5\times10^9~\mathrm{[K]})$.
Figure \ref{fig:T_cntr} shows that the remaining time between the central carbon depletion phase and the core collapse is \old{orders of} $\sim 10\,{\rm years}$. Our discussion in the main text \old{focused} on the stellar properties with $t_{\rm lb}\gg 10\,{\rm years}$, and this indicates that it is reasonable to approximate the central carbon depletion phase as the moment of the core collapse.
\old{We also note that both the lifetime of the star and the central temperature at the moment of the central carbon depletion stage are consistent with the properties suggested in \citet{2010ApJ...718..357T}.}

\old{We have defined the mass fraction of carbon at the moment of entering the carbon depletion stage as $10^{-6}$. It corresponds to the stage immediately before oxygen burning stage in the core. Since the timescale of the oxygen burning stage would not be largely affected by stellar properties such as stellar mass, we expect that our criterion of the carbon depletion stage is robustly tracing the epoch immediately before the core collapse, and the assumption of $t_{\rm end}-t_{\rm cc}=10\,{\rm years}$ is reasonable.}

\old{On the other hand, we find that there is a variation in the definition of the carbon depletion stage depending on the paper. \citet{2020A&A...637A...6L} defines carbon depletion as the moment when the mass fraction of the carbon in the central core decreases below $10^{-4}$. This results in the initiation of the carbon depletion $100\,{\rm years}$ before the core collapse. \citet{2023arXiv230801819E} declares carbon depletion $10^3\,{\rm years}$ before the core collapse, and we expect this is due to the further higher criteria for the central carbon mass fraction.}

\section{Comparison between the analytic and numerical solution of the CSM structure}\label{app:CSM}

While we performed the numerical simulations of the CSM formation in Section \ref{subsec:csm}, it is possible to rely on the analytical description of the wind structure. \cite{2020ApJ...894....2P} presented the framework describing the hydrodynamical structure of the CSM originating from the time-dependent mass-loss history. Comparing these two methods allows us to confirm the application limit of the analytical treatment in the context of CSM formation.

\begin{figure}
    \centering\includegraphics[width=0.49\textwidth]{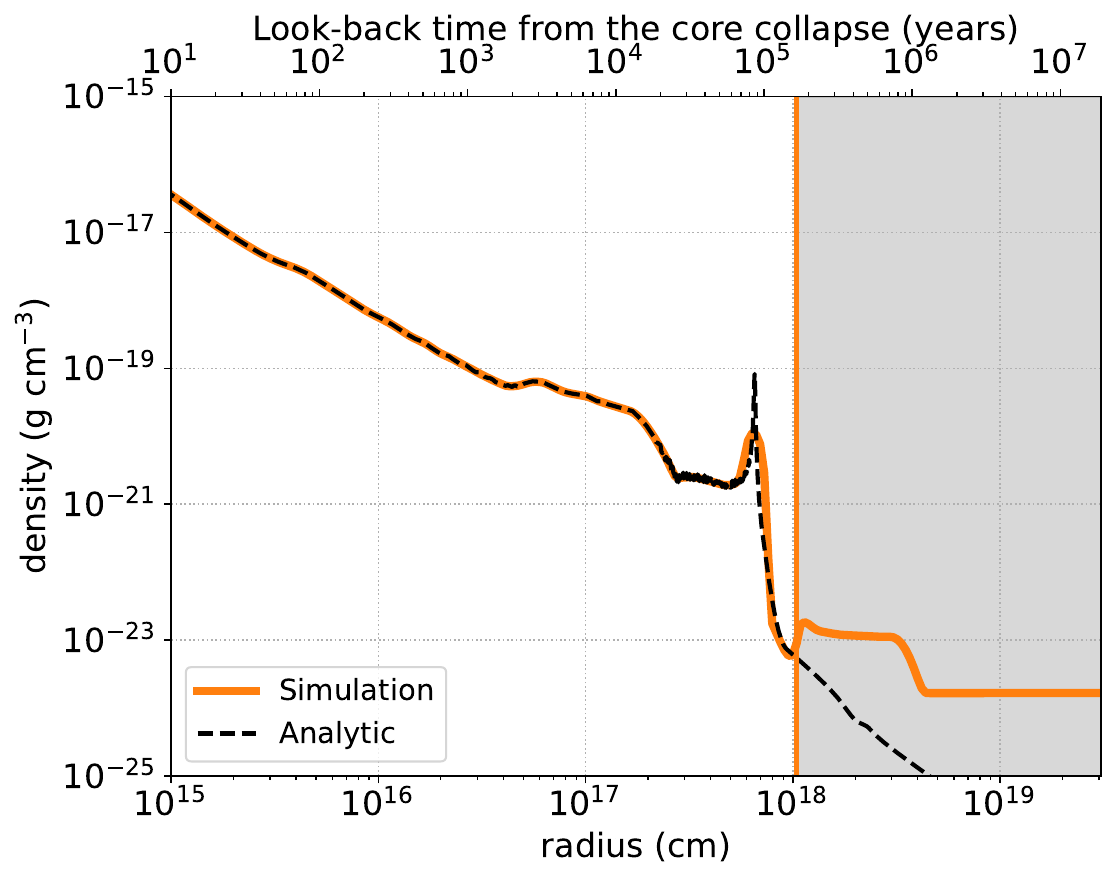}
    \caption{Comparison of the reconstructed CSM density structure between numerical (orange solid line) and analytic (black dashed line) solutions in the model with $M_1=12\,M_\odot, M_2=10.8\,M_\odot, P_{\rm orb}=1300\,{\rm days}$. The gray-shaded region on the right side indicates the deviation of the analytical solution from the numerical one.}
    \label{fig:CSM_ananum}
\end{figure}

Figure \ref{fig:CSM_ananum} shows the comparison between our numerical model and the analytical treatment with the binary parameter of $M_1=12\,M_\odot, M_2=10.8\,M_\odot, P_{\rm orb}=1300\,{\rm days}$. It is observed that these two treatments are compatible with each other within $r\sim 10^{18}\,{\rm cm}$.
The corresponding look-back time of this radius is $\sim 0.1\,{\rm Myr}$, indicating that the time variation of the mass-loss rate after the helium exhaustion in the core can be directly associated with the density variation of the CSM.
In this model, actually, the mass-loss rate is greatly enhanced immediately after the helium exhaustion, and this feature is clearly appearing as the shell-like structure around several times $10^{17}\,{\rm cm}$.
We confirm the validity of the analytical treatment of time-dependent wind propagation in \cite{2020ApJ...894....2P}.
\old{In the outer region from $\sim 10^{18}\,{\rm cm}$, the thermal pressure of the interstellar medium heated by the wind's shock prevents the further expansion of the stellar wind. Then the heated interstellar medium and the stellar wind would accumulate at the contact discontinuity, and the resultant CSM structure deviates from the analytical solution \citep[see also][]{1977ApJ...218..377W, 2022ApJ...930..143M}.}

\section{An interpretation of the radial expansion of the primary star suffering from mass removal}\label{app:M_R}

\tm{A star suffering from the mass reduction expands its radius because of the increase in the gravitational energy compared to the internal energy, as explained in Section \ref{subsec:binary}. In this section, we provide quantitative analysis on the response of the stellar radius to mass reduction on the basis of the polytrope sphere of gas.}

\tm{Firstly let us consider a polytrope sphere of gas in hydrostatic equilibrium with the polytropic index $n$ before the mass variation.
As is known, we can solve the radial structure of the density of the polytrope sphere from equation of hydrostatic equilibrium and the Poisson equation \citep{1990sse..book.....K}.
We can also write down the gravitational energy of the system as 
\begin{eqnarray}
E_{\rm grav}= -\frac{3}{5-n}\frac{GM^2}{R},
\end{eqnarray}
where $M$ and $R$ are the mass and radius of the polytrope sphere, respectively.
On the ground of the Virial theorem, the total energy of the gravitationally bounded system can be associated with the gravitational energy as follows: \citep{1990sse..book.....K},
\begin{eqnarray}
E_{\rm tot} = \frac{3\gamma-4}{3\gamma-3}E_{\rm grav} = -\left(\frac{3\gamma-4}{3\gamma-3}\right)\frac{3}{5-n}\frac{GM^2}{R} \equiv -F(\gamma,n)\frac{GM^2}{R},
\end{eqnarray}
where $\gamma$ is the specific heat ratio of the gas. For convenience, the coefficients relevant to $\gamma$ and $n$ are all absorbed in $F(\gamma, n)$, which we have newly introduced.
}

\tm{Next, we consider that a perturbation from an external force induces the stellar mass modification of $\delta M$, corresponding to the variation in the gravitational energy of the star $\delta E$ and in the stellar radius of $\delta R$, but the system is still gravitationally bounded. For now we do not fix the sign of $\delta M$, $\delta R$, and $\delta E$. Then, by definition,
\begin{eqnarray}
\delta E = -\int_R^{\infty} \frac{GM \delta M}{r^2} dr = -\frac{GM}{R}\delta M
\end{eqnarray}
is satisfied. The relationship among the perturbation terms is given as
\begin{eqnarray}
\frac{\delta E}{E_{\rm tot}} = 2\frac{\delta M}{M} - \frac{\delta R}{R}.
\end{eqnarray}
Substituting the formulae of $E_{\rm tot}$ and $\delta E$ into this equation, we can find out
\begin{eqnarray}
\frac{\delta R}{R} = \left(2-\frac{1}{F(\gamma,n)}\right)\frac{\delta M}{M}.\label{eq:deltaR}
\end{eqnarray}
Note that when we consider mass reduction from the polytrope sphere, $\delta M/M$ is negative.}

\begin{figure}
    \centering
    \includegraphics[width=0.5\linewidth]{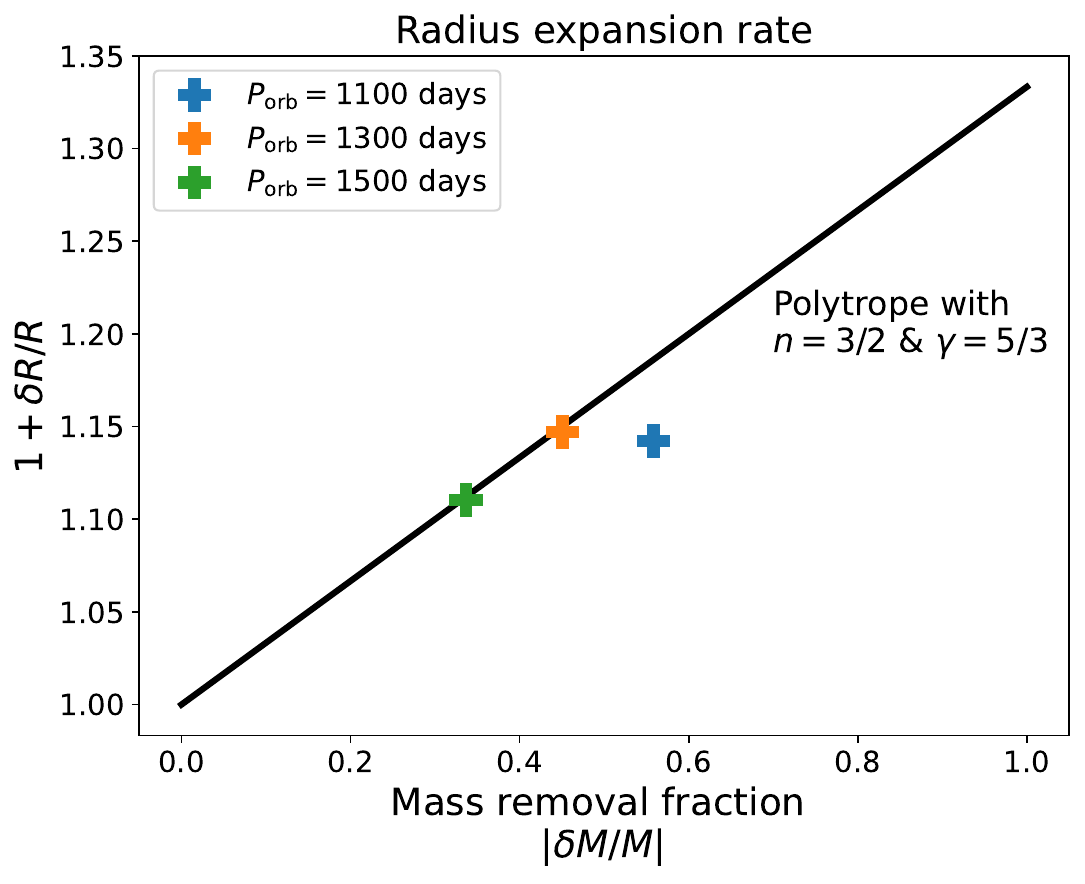}
    \caption{\tm{Radial expansion rates with response to the perturbation ($1+\delta R/R$) as a function of the mass removal fraction $|\delta M/M|$ is illustrated on the assumption of $\gamma=5/3$. Equation (\ref{eq:deltaR}) with $n=3/2$ is plotted by the black solid line, while the numerical values obtained from our binary models in Section \ref{sec:result} are plotted by the thick cross points.}}
    \label{fig:coeff_n}
\end{figure}

\tm{Equation (\ref{eq:deltaR}) clarifies the response of the stellar radius to the stellar mass variation. Figure \ref{fig:coeff_n} shows the radial expansion rate ($1+\delta R/R$) as a function of the mass removal fraction $|\delta M/M|$ on the assumption of $\gamma=5/3$. We plot the relationship with the polytrope index $n=3/2$, which is plausible for the modeling for a star engulfed by a convective envelope \citep[e.g,.][]{1999ApJ...510..379M}.
We also plot the numerical data obtained from our binary simulations in Section \ref{subsec:binary} to compare with the polytrope analysis.
We can see that our binary models with $P_{\rm orb}=1300$ and $1500\,{\rm days}$ undergo the mass removal of $45$ and $33\%$ of the initial mass in total, and result in $15$ and $11\%$ of the radial expansion, respectively. These variations are in good agreement with the predictions from the analysis based on the polytrope sphere, which is illustrated by the black line.
As for the model with $P_{\rm orb}=1100\,{\rm days}$, the expansion rate is suppressed to $14\%$ while predicted as $18\%$.
Anyway, our analysis based on the polytrope sphere can successfully explain $\sim 10\%$ of the radial expansions of the primary stars that suffer from the mass stripping in the binary.
We expect that the precise quantification of the stellar radius should depend on the physics we have neglected, but that the inflation of ten percents of the stellar radius seen in our binary models can be robustly ascribed by the stripping of the envelope through binary interaction.}

\bibliography{manuscript}{}
\bibliographystyle{aasjournal}

\end{document}